# Symmetry of the carbon nanotube modes and their origin from the phonon branches of graphene

M. V. Avramenko[*], S. B. Rochal, Yu. I. Yuzyuk

*Department of Nanotechnology, Faculty of Physics, Southern Federal University, 5, Zorge str., Rostov-on-Don, 344090, Russia*

Dated: 26 July 2012

*avramenko.marina@gmail.com

A new group-theory method to relate a spectrum of carbon nanotubes with that of a graphene monolayer is elaborated. The spectrum reconstruction is performed using a virtual intermediate planar periodic structure. Selected irreducible representations of its space symmetry group and all those of the nanotube one are isomorphic and therefore span the correlated excitations. The method is applied to study the origin of the zone-center phonon modes of chiral and achiral carbon nanotubes. The structure of the G-band for particular types of nanotubes is determined. The results obtained could be useful for experimental indexing of the carbon nanotubes by means of Raman spectroscopy.

**Keywords:** single-walled carbon nanotubes, graphene, Raman spectroscopy, zone-folding theory, factor-group analysis, phonon modes

A graphene layer (GL) possesses a higher symmetry with respect to a single wall carbon nanotube (SWCNT) and therefore, has relatively simple phonon dispersion branches structure [1, 2]. Comprehensive understanding of general correlations between phonons of GL and those of the SWCNT is obviously required. The Brillouin zone folding scheme, explaining formation of the one-dimensional reciprocal space of SWCNTs from the two-dimensional space of the GL was developed more than 10 years ago [3]. This fruitful theory explained electronic zone structure of the SWCNT and the criterion for classification of SWCNTs as metallic or semiconducting ones was elaborated [3]. Besides, the factor-group analysis of SWCNTs vibrational spectra was carried out as well, and the lists of Raman- and IR-active modes (in both chiral and achiral cases) were found [4, 5]. Later, it was demonstrated that all SWCNTs irrespective to their chirality possess the higher symmetry than it had been considered before and the factor-group analysis of SWCNTs vibrational spectra was revised [6]. Since the higher symmetry imposed more rigorous selection rules the number of Raman- and IR-active modes reported in [6] was smaller in comparison with earlier results [4,5]. Although the authors of Refs. 3-5 admitted [7] the results of [6], the revision of the GL reciprocal space folding theory still has not been carried out (see, for example, the latest review [9]). In our opinion, an application of the initial theory [3] leads to the correct results in those cases only, for which the exact symmetry of SWCNT vibrational or electronic states is not very important. The earlier theory [3] describes correctly the origin, number and polarization of the Raman-active vibrational modes in the case of the lowest-symmetry chiral SWCNTs, but it is not so for higher-symmetry achiral tubes. However, an approach, considering the origin of all IR- and Raman-active modes in achiral SWCNTs and consisting with the correct symmetry [6, 12] is not known till now.

The aim of this Letter is to develop a general approach describing both the origin and symmetry of vibrational modes in the SWCNTs of all possible types. We pay a particular attention to the origin of Raman-active modes of achiral SWCNTs. The obtained results will be very useful for experimental identification of SWCNT by means of Raman spectroscopy, because number of G-band components is shown to be different for chiral and achiral SWCNT. This method could have experimental applications concerning the SWCNT type determination, the tube indexing and the study of the modes, whose frequencies are strongly dependent on the tube indices.

First of all, let us emphasize an essential difference between the phase transitions in crystals with the unit cell enlargement and the virtual rolling procedure transforming a GL into a SWCNT. In contrast to a conventional transition with the unit cell enlargement in crystals, formation of the SWCNT from the graphene sheet never makes the carbon positions symmetrically nonequivalent. The translational symmetry of the GL does not vanish after rolling, but transforms into the rotational one within the SWCNT. The primitive GL translations $a_1$ and $a_2$ of the hexagonal lattice become the screw rotations about the principal axis of the tube with the perimeter translation $P = na_1+ma_2$, where $n$ and $m$ are two integer indices. The angles corresponding to these rotations are given by

$$\varphi_{a_1} = 2\pi \frac{2n+m}{2(n^2+m^2+nm)}, \varphi_{a_2} = 2\pi \frac{2m+n}{2(n^2+m^2+nm)} \tag{1}$$

The simultaneous existence of rotations (1) leads to the appearance of the $N$-fold screw axis $C_N^z$ with

$$N = \frac{2(n^2+m^2+nm)}{GCD(2n+m,2m+n)} \tag{2}$$

where GCD $(2n+m, 2m+n)$ is the greatest common devisor of two integers $(2n+m)$ and $(2m+n)$. As known [3], $N$ is always integer, and is equal to the number of hexagons contained within the nanotube unit cell. Moreover, the value $N$ is always even [4]. The SWCNTs have a set of the two-fold axes $U_2^{(i)}$ which are perpendicular to the principal $C_N^z$ axis [3]. Let us clarify the origin of these axes. According to the well-known crystallographic theorem, any two two-fold axes $C_2^{(j)}$ and $C_2^{(k)}$ of the unrolled GL are always spaced by the vector $P/2$. When the graphene layer is rolled up these $C_2^{(j)}$ and $C_2^{(k)}$ axes form one $U_2^{(i)}$ two-fold axis of the SWCNT. Since the $C_2^{(m)}$ axes cross either the middles of hexagons or the middles of their edges, their total number is equal to $4N$ per the unrolled nanotube unit cell and is twice smaller when the SWCNT in rolled up. The factorization of the nanotube space symmetry with respect to the translational subgroup reduces this number to $N$. In other words, the factor-group of any SWCNT always contains $N$ $U_2^{(i)}$ axes [3].

Let us call the GL translation corresponding to the rotation angle $2\pi/N$ around the $C_N^z$ axis as characteristic translation $a_\varphi$. Obviously, its projection on $P$ is $N$ times smaller than $|P|$. Note, that other translations which have the projection smaller than one of $a_\varphi$ do not exist.

The two basis vectors of the extended SWCNT cell in reciprocal space are $b_\varphi$ and $b_z$ [3]. The first of them, $b_\varphi$, is parallel to $P$ and is $2\pi/|P|$ in length, and $b_z$ is perpendicular to the first one. They are expressed as follows:

$$b_\varphi = \frac{(2n+m)b_1 + (2m+n)b_2}{2(n^2+m^2+nm)}, \tag{3}$$

$$b_z = \frac{mb_1 - nb_2}{N}, \tag{4}$$

where $b_1$ and $b_2$ are the basis vectors of GL reciprocal lattice [3-5]. Another important relation is given by

$$a_\varphi b_\varphi = \frac{2\pi}{N} \tag{5}$$

After the nanotube rolling up, any point in the reciprocal space of the GL with coordinates
$$\mu \boldsymbol{b}_\varphi + (q + \rho)\boldsymbol{b}_z \qquad (6)$$

where $\mu$ and $\rho$ are integers and $|q|<|\boldsymbol{b}_z|/2$, coincides with the point $q\boldsymbol{b}_z$ of the one-dimensional first Brillouin zone of SWCNT. Expression (6) defines a corresponding set of cutting lines [5], which are indexed by two integer indices $\mu$ and $\rho$. The equivalent lines in this set are related by the translations of the GL reciprocal lattice. In order to find the correct correlation between the vibrational modes of GL and SWCNT we have to consider nonequivalent lines only. The number of nonequivalent nodes and corresponding cutting lines is equal to the multiplication $N$ of the unit cell in the direct space [5]. Nonequivalent lines can be easily excluded assuming $\rho=0$ in Eq. (6) and considering all nodes within the strip with the width equal to $|\boldsymbol{b}_z|$. Since the denominator in (3) is equal to $N\times GCD(2n+m, 2m+n)$, the nodes and corresponding lines with indices $(\mu, 0)$ and $(\mu+N, 0)$ are equivalent, and no other equivalent lines within the period $N\boldsymbol{b}_\varphi$ can be found. So this period defines the length of rectangle containing all the nonequivalent lines, and allowed $\mu$ values are in the interval $-N/2 <\mu\leq N/2$. The nonequivalent lines were chosen in a similar way in the pioneer work [3].

Symmetry lowering caused by the rolling up a GL into a SWCNT can be divided into two steps. At the first step the translational symmetry of GL is preserved but orientational symmetry lowers from $D_{6h}$ to chiral $C_2$ or achiral $C_{2v}$ group depending on the nanotube axis direction. We denote the GL with the lowered symmetry as GLLS and first step (GL→GLLS) leads to the reduction of irreducible representation (IRREPs) of the GL space symmetry group according to the conventional group-theory methods [8].

At the second step we replace the two-dimensional (2D) translational symmetry of GLLS by the screw rotational symmetry of the SWCNT. This symmetry lowering is homomorphic. Any two translations $t_1$ and $t_2$ of the GLLS become equivalent in the nanotube, provided $t_1$-$t_2$ = $\boldsymbol{P}$. In spite of the above homomorphism, IRREPs of GLLS corresponding to the reciprocal space points defined by (6) and IRREPs of the SWCNT are isomorphic, because these IRREPs of GLLS have a homomorphism kernel containing the translation $\boldsymbol{P}$. Obviously, the other GL IRREPs which are not related to the cutting lines defined by (6) are not isomorphic to the nanotube's IRREPs.

Below we apply this approach to assign all vibrational modes of SWCNT with the wave vector $\boldsymbol{q}=0$. According to our previous consideration, these modes originate from the centers of cutting lines, namely from the equidistant points with the coordinates $\mu\boldsymbol{b}_\varphi$, where $-N/2 <\mu \leq N/2$. For these modes the index $\mu$ has a simple physical sense: $|\mu|=|\boldsymbol{P}|/\lambda$, where $\lambda$ is the wavelength of the corresponding mode. Since $N=2n$ for the achiral zigzag ($\boldsymbol{P}=n\boldsymbol{a_2}$) and armchair ($\boldsymbol{P} = n(2\boldsymbol{a_1}-\boldsymbol{a_2})$) nanotubes, the coordinates $\mu\boldsymbol{b}_\varphi$ are simplified to the following forms
$$\mu\frac{2\boldsymbol{b}_2 + \boldsymbol{b}_1}{N} \text{ and } \mu\frac{\boldsymbol{b}_1}{N}. \qquad (7)$$

Note that the first and second vector sets defined by Eq. (7) are directed to the K and M points of the first Brillouin zone, respectively (see fig. 1).

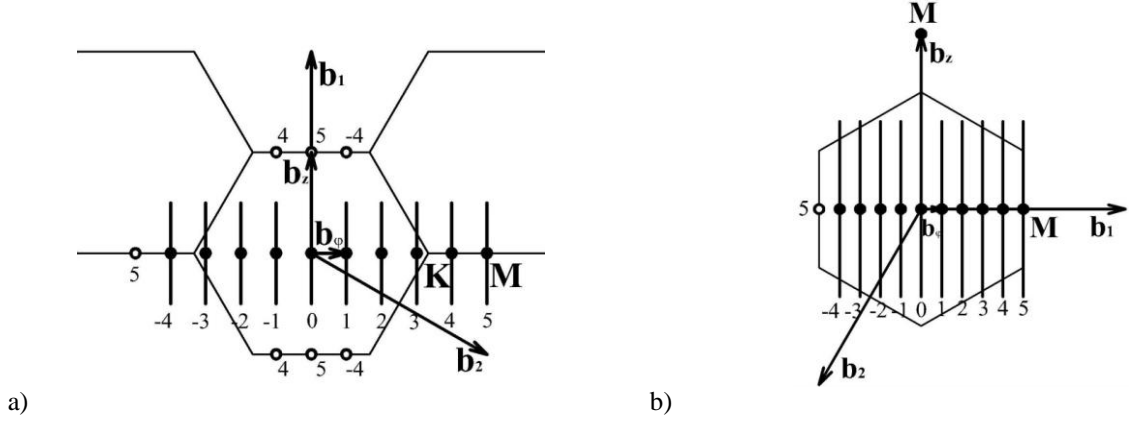

a)  b)

Figure 1. Formation of the first Brillouin zone of the achiral SWCNTs from the nonequivalent lines cutting the reciprocal space of the GL. a) zigzag ($P=5a_2$); b) armchair ($P=10a_1-5a_2$). In both cases $N=10$ and the number of equally spaced cutting lines is also 10. The nodes of the reciprocal lattice with the basis vectors defined by (3-4) are denoted by circles. Nodes with the same numbers are equivalent with respect to the GL reciprocal lattice translations. Analogous figures for chiral SWCNT can be found in [9].

Functions $\phi_k^j = u_k^j \exp(i\mathbf{k}\mathbf{r})$ (where $\mathbf{r}$ is a radius vector and $u_k^j$ are the periodic functions, spanned by IRREPs of the point group $G_\mathbf{k}$ of the wave vector $\mathbf{k}$) form the basis for IRREPs of any symmorphic space group [11] (including the GLLS space group). Irrespective to the symmetry of GLLS the star of wave vector $\mathbf{k}= \mu \mathbf{b}_\varphi$ has one or two arms and the group $G_\mathbf{k}$ has 1D IRREPs only. Therefore, the IRREPs of the GLLS space group as well the $q=0$ IRREPs of the SWCNT group are one- or two-dimensional. One-dimensional IRREPs originate from the points, where vectors $\mathbf{k}$ and $-\mathbf{k}$ are equivalent (in other words, the wave vector star has one arm). These are the points with $\mu=0$ and $\mu=N/2$, corresponding to Γ and M points of the Brillouin zone of the GL, respectively. Note, that $\mu=N/2$ corresponds to M point for both chiral and achiral tubes. For achiral SWCNTs this fact can be also shown by substituting $\mu=N/2$ in (7). Indeed, the points $\mu=0$ and $\mu=N$ are the translationally equivalent nodes of the GL reciprocal lattice. The point with $\mu=N/2$ is located exactly between these nodes (but does not coincide with another node of the GL reciprocal lattice) and, therefore, lies on the two-fold axis and corresponds to the M point of Brillouin zone of the GL with the wave vector possessing the three-arm star $\left\{\frac{1}{2}\mathbf{b}_1, \frac{1}{2}\mathbf{b}_2, \frac{1}{2}(\mathbf{b}_1+\mathbf{b}_2)\right\}$. During the first step (GL→GLLS) of the symmetry lowering this star is split and only one of its arms fall into the Brillouin zone center of SWCNT.

Thus all vibrational $q=0$ modes of SWCNTs originate from the two high-symmetry points (Γ and M) and from ($N$-2) low-symmetry points of the Brillouin zone of the GL. Now we can proceed to the correlation analysis of the IRREPs spanning these modes. Since the first step of the symmetry lowering is quite conventional, it is not considered here in details. During the second step we can identify the isomorphism between the IRREPs of the GLLS and SWCNTs using the generator method [see Appendix A]. According to this approach, two IRREPs are isomorphic if, and only if these two IRREPS can be presented in the forms with identical generator matrices. Let us start from the high-symmetry points associated with 1D IRREPs. As it has been mentioned above, the GL characteristic translation $\mathbf{a}_\varphi$ is transformed into the screw rotation $C_N^1$ in the SWCNT, so according to [1110], the matrixes of $g(C_N^1)$ and one-dimensional translation $\mathbf{a}_\varphi$ are the same: $g(C_N^1)=g(\mathbf{a}_\varphi)=\exp(ika_\varphi)$. Using (5), we obtain that 1D IRREPs of types A ($g=1$) and B ($g=-1$) originate from the points Γ and M, respectively. The symmetry of chiral SWCNTs is described by $D_N$ groups generated by the $N$-fold rotation $C_N^1$ and 2-fold rotation $U_2$ [see Appendix A]. The symmetry generator $U_2$ is common for GLLS and nanotube and its matrix $\mathbf{g}(U_2)$ is equal to 1 or −1, that distinguishes between IRREPs with subscripts "1" and "2". Thus, the 1D IRREPs correlation for chiral nanotubes is determined (see columns 1, 2 and 3 in Tables 1 and 2).

The symmetry of achiral SWCNTs is described by the $D_{Nh}$ groups. To work with this group besides the above two generators the third one is required. We prefer to use a mirror plane $\sigma_h$ perpendicular to the principal axis as this generator, since the plane $\sigma_h$ always belongs to the $G_k$ group in the achiral GLLS. Note, that there are two types of $D_{Nh}$ IRREPs: even (subscript "g") and odd (subscript "u") with respect to inversion. The type of IRREP can be easily defined using of the following relation for the inversion matrix: $\mathbf{g}(I)=(\mathbf{g}(C_N^1))^{N/2}\mathbf{g}(\sigma_h)$. This formula allows us to obtain the correlation GL→GLLS→SWCNT of 1D vibrational modes of achiral SWCNTs. (See columns 1, 4, 5 and 6 in Tables 1 and 2). The parallel and perpendicular polarization directions with respect to the GL are denoted by letters 'i' and 'o' in the first columns of both Tables 1 and 2.

Table 1. The vibrational modes correlation GL→GLLS→SWCNT in the Γ point. The IRREPs of the layer space groups (columns 1, 2 and 4) are denoted by the same symbols as the IRREPs of the corresponding factor groups, which are $D_{6h}$, $C_2$ and $C_{2v}$, respectively.

| GL | Chiral GLLS | Chiral SWCNT | Achiral GLLS | Zigzag SWCNT | Armch.SWCNT |
|---|---|---|---|---|---|
| $E_{2g}$ (iTO and iLO) | $2A$ | $2A_1$ | $A_1+A_2$ | $A_{1g}+A_{1u}$ | $A_{1g}+A_{1u}$ |
| $B_{1g}$ (oTO) | $B$ | $A_2$ | $B_1$ | $A_{2u}$ | $A_{2g}$ |
| $E_{1u}$ (iTA and iLA) | $2B$ | $2A_2$ | $B_1+B_2$ | $A_{2u}+A_{2g}$ | $A_{2g}+A_{2u}$ |
| $A_{2u}$ (oTA) | $A$ | $A_1$ | $A_1$ | $A_{1g}$ | $A_{1g}$ |

Table 2. The vibrational modes correlation GL→ GLLS →SWCNT in the M point. The IRREPs of the layer space groups (columns 1, 2 and 4) are denoted by the same symbols as the IRREPs of the corresponding $G_k$ groups, which n the M point are $D_{2h}$, $C_2$ and $C_{2v}$, respectively. Dotted vertical lines divide the cells of the 5-th and 6-th columns into two parts. The left and rights parts of each columns correspond to the cases $N=4m+2$ and $N=4m$, respectively, where $m$ is a positive integer. This correction of results [6] was found in [12].

| GL | Chiral GLLS | Chiral SWCNT | Achiral GLLS | Zigzag SWCNT | | Armch.SWCNT | |
|---|---|---|---|---|---|---|---|
| $A_{1g}$ (iTO) | $A$ | $B_1$ | $A_1$ | $B_{1u}$ | $B_{1g}$ | $B_{1u}$ | $B_{1g}$ |
| $A_{2g}$ (iLO) | $A$ | $B_1$ | $A_2$ | $B_{1g}$ | $B_{1u}$ | $B_{1g}$ | $B_{1u}$ |
| $B_{2g}$ (oTO) | $B$ | $B_2$ | $B_1$ | $B_{2g}$ | $B_{2u}$ | $B_{2u}$ | $B_{2g}$ |
| $B_{1u}$ (iTA) | $B$ | $B_2$ | $B_1$ | $B_{2g}$ | $B_{2u}$ | $B_{2u}$ | $B_{2g}$ |
| $B_{2u}$ (iLA) | $B$ | $B_2$ | $B_2$ | $B_{2u}$ | $B_{2g}$ | $B_{2g}$ | $B_{2u}$ |
| $A_{2u}$ (oTA) | $A$ | $B_1$ | $A_1$ | $B_{1u}$ | $B_{1g}$ | $B_{1u}$ | $B_{1g}$ |

And now let us focus on the remaining $q=0$ vibrational modes associated with integer $\mu$ values originated from the interval $0<|\mu|<N/2$. For this interval the star of the wave vector $\mathbf{k}=\mu\mathbf{b}_\varphi$ has six arms in the high-symmetry GL. After the first step of the symmetry lowering GL

→GLLS this star is reduced to two arms: $\mu\boldsymbol{b}_\varphi$ and $-\mu\boldsymbol{b}_\varphi$. For the chiral GLLS the $G_k$ group is trivial and for each $\mu$ value there only one IRREP exists. The $G_k$ group corresponding to the achiral GLLS is generated by the mirror plane $\sigma_h$ and contains two symmetry elements only. Accordingly, for each $\mu$ value there are two non-equivalent IRREPs. In both chiral and achiral cases the 2D IRREP of the nanotube symmetry group originates simultaneously from two points $\mu\boldsymbol{b}_\varphi$ and $-\mu\boldsymbol{b}_\varphi$. The isomorphism between the GLLS and nanotube IRREPs is established using our generator approach. The matrix $\boldsymbol{g}(C_N^1)$ coincides obviously with the matrix $\boldsymbol{g}(a_\varphi)$ (see Table 4 of the Appendix A). The second generator of the nanotube symmetry is the two-fold axis $U_2$. (See the explicit form of its matrix in the same Table). Let choose the coordinate origin in the GLLS to be coincided with one of the $C_2$ axes transforming into the $U_2$ axis of the nanotube. Than the matrix $C_2$ permutes simply the IRREP basis functions and the both matrices $\boldsymbol{g}(C_2)$ and $\boldsymbol{g}(U_2)$ always coincide. Since the GLLS (like the GL) has 6 phonon branches, the number of the corresponding $\boldsymbol{q}=0$ vibrational modes ($0<|\mu|<N/2$) is equal to $6(N/2-1)$. In the chiral case these modes are spanned by the $E_\mu$ IRREP. In the achiral case the further identification is required to distinguish between $E_{\mu u}$ and $E_{\mu g}$ modes.

The existence of the $E_{\mu u}$ and $E_{\mu g}$ IRREPs in the achiral case is due to the fact that the achiral GLSS has two non-equivalent IRREPs corresponding to the one $|\mu|$ value. The basis functions of the first IRREP are even with respect to the mirror plane $\sigma_h$, while the basis functions of the second one are odd. Therefore, for these IRREPs matrix $\boldsymbol{g}(\sigma_h)$ is the identity and minus identity matrix, respectively. Note also that if $\mu$ is even, the matrix for two-fold axis $(\boldsymbol{g}(C_N^1))^{N/2}$ is the identity one. So the even/odd parity with respect to the $\sigma_h$ plane means the even/odd parity with respect to inversion transformation of the nanotube. If $\mu$ is odd, the rule is opposite. Table 3 shows the parity of the GL modes with respect to the $\sigma_h$ plane. Of course, these parities are not changed after the rolling procedure.

Table 3. The parities of the GL modes with respect to $\sigma_h$. We use the conventional mode notations for the GL where letters "i" and "o" (in plane, out plane) denote polarization direction with respect to the mode wave vector $\mu\boldsymbol{b}_\varphi$.

|  | iLA | iTA | oTA | iLO | iTO | oTO |
|---|---|---|---|---|---|---|
| Armchair | 1 | -1 | 1 | 1 | -1 | 1 |
| Zigzag | 1 | -1 | 1 | -1 | 1 | -1 |

Using the above obtained results we can determine the Raman-active modes of the achiral SWCNTs, originating from the optical vibrations of GL. It is important to discuss the difference between the structure of so-called G-band (originated from optical iLO and iTO modes of GL) in chiral and achiral SWCNTs.

In achiral SWCNTs all the GLLS modes, which are odd with respect to $\sigma_h$, in the point $\mu=1$ possess the $E_{1g}$ symmetry, while all the even GLLS modes in the point $\mu=2$ belongs to $E_{2g}$ type symmetry. Therefore G-band of achiral SWCNTs consists of three modes: one $A_{1g}$, (see Table 1), one $E_{1g}$ and one $E_{2g}$. In armchair SWCNTs $E_{1g}$ and $E_{2g}$ modes originate from iTO and iLO mode of GL, respectively. Let us stress, that the origin of these modes in zigzag SWCNTs is opposite (see Fig. 2).

The G-band structure for the chiral SWCNT was described correctly in previous works (see [9], for example). The selection rules based on the mode parity with respect to the reflection $\sigma_h$ are irrelevant for the chiral symmetry, and each acoustic and optic phonon dispersion curve of the GL generates one $E_1$ and one $E_2$ Raman-active modes. Accordingly, the G-band consists of 6 modes: two split $A_1$ modes, two $E_1$ and two $E_2$ modes. Also note, that in Raman spectra of chiral SWCNTs the frequencies of two modes originated from oTO phonon dispersion branch (points $\mu=1$ and $\mu=2$) should be close. But in the achiral case this curve gives the origin to the single mode, which has the $E_{1g}$ and $E_{2g}$ symmetry in zigzag and armchair nanotubes, respectively. We hope that this peculiarity and the quite different structure of the G-band in the SWCNTs of different types can be useful for experimental identification of the nanotubes types. We therefore expect that these optical modes can be observed separately by polarization-dependent Raman measurements.

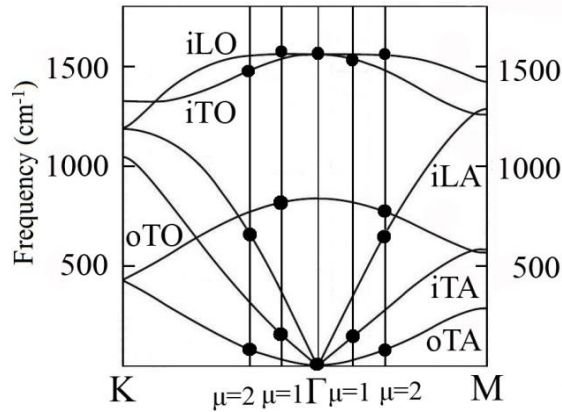

Figure 2. The origin of Raman-active modes in the zigzag and armchair SWCNTs is shown on the left and right sides with respect to the $\Gamma$ point, respectively. Filled circles correspond to the Raman-active modes. Phonon dispersion curves of the GL are adapted from [13].

The proposed methodology and selections rules can be easily applied to identify Raman-active modes originating from acoustical branches of the GL. Frequencies of the vibrational modes of SWCNT, originating from the acoustical ones strongly depends on $\boldsymbol{b}_\varphi$ value, which is in turn defined by $n$ and $m$ indices. In addition to well-known SWCNT assignments based on the RBM frequency and Kataura-plot [6], a new approach for SWCNT identification can be actually developed using our results. However, it is worth to note, that wave numbers of low-frequency modes originating from acoustical branches with small wave vectors ($\boldsymbol{k} \approx 0$) should be very sensitive to the curvature and may exhibit significant shifts with going from the flat GL to SWCNT.

The rolling procedure can additionally (with respect to the zone folding approach) shift the frequencies of these modes, that makes more complex an experimental application of our results. This frequency deviation is especially essential for the oTA mode in the region of small wave vectors, where the Goldstone behavior of this mode is transferred from the point $\mu=0$ (in the GL) to the point $\mu=1$ (in SWCNTs). Thus, the bending mode of the GL with zero wave vector and frequency is transformed into the nanotube radial breathing mode, which frequency is usually explored for the nanotube indexing [14]. The essential frequency shift of the modes originated from the oTA mode can be explained qualitatively in the following way. On the one hand any bending mode essentially parameterizes (much stronger than the other modes) the shape of the surface which it propagates along, and on the other hand, the rolling procedure also changes the surface shape remarkably. The coupling between these phenomena is resulted in the essential frequency shift of the oTA modes with small wave vector (and small $\mu$ values). In contrast, the iTA and iLA modes are not so sensitive to the rolling and remain of the Goldstone type in the $\mu=0$ point. Thus the associated frequency shift should be weaker, and possibly, the frequency analysis of the nanotube modes originating from iTA and iLA dispersion curves could provide a new experimental method for the SWCNT indexing. In any case a quantitative description of acoustic modes frequency shifts resulted from the topology change could be a subject for further experimental and theoretical studies.

In conclusion, we have demonstrated the restrictions of the pioneer theory [3-5] and developed a general approach comprehensively explaining the origin of the vibrational modes of carbon nanotubes from the phonon branches of the graphene monolayer. The method is applied to identify the zone-center optical phonon modes corresponding to the G-band of chiral and achiral carbon nanotubes. The low-frequency modes originating from the acoustic branches are also discussed.

# Appendix A

## SWCNTs symmetry groups, their irreducible representations

$D_N$ groups with even $N$ are generated by 2 symmetry elements: the $N$-fold rotation $C_N$ and 2-fold rotation $U_2$ (the latter axis is perpendicular to the principal one). The existence of $C_N$ axis implies $N$ $U_2$ rotations, and they form 2 classes of conjugate elements with $N/2$ axes per each class. Each of $D_N$ group has 4 one-dimensional representations and $N/2-1$ two-dimensional ones (denoted as $E_\mu$, $\mu=1, 2\ldots N/2-1$). Matrices of all irreducible representation generators are listed in Table 4.

Table 4. Matrices of irreducible representations corresponding to generators of $D_N$ symmetry group Note, that functions $\varphi_1^\mu = (x-iy)^\mu$ and $\varphi_2^\mu = (x+iy)^\mu$ are spanned by the matrixes of the $E_\mu$ representation presented in the last row of this Table.

| $D_N$ | $g(C_N)$ | $g(U_2)$ |
|---|---|---|
| $A_1$ | +1 | +1 |
| $A_2$ | +1 | -1 |
| $B_1$ | -1 | +1 |
| $B_2$ | -1 | -1 |
| $E_\mu$ | $\begin{bmatrix} e^{\frac{2\pi i}{N}\mu} & 0 \\ 0 & e^{-\frac{2\pi i}{N}\mu} \end{bmatrix}$ | $\begin{bmatrix} 0 & 1 \\ 1 & 0 \end{bmatrix}$ |

$D_{Nh}$ groups with even $N$ are generated by 3 symmetry elements: 2 generators of $D_N$ groups and the mirror plane $\sigma_h$, which is perpendicular to the principal axis. Thus, the number of elements and the number or IRREPs is twice larger with respect to $D_N$ groups [15]. IRREPs become even (subscript "$g$") or odd (subscript "$u$") with respect to inversion. Inversion matrix is identity or minus identity for "$g$" and "$u$" IRREPs, respectively.

Different symmetry elements can be expressed using the generators according to the following
1) The matrix for 2-fold rotation about the principal axis: $g(C_2)=(g(C_N^1))^{N/2}$;
2) The matrix for mirror plane $\sigma_h$, which is perpendicular to the principal axis: $g(\sigma_h)=(g(C_N^1))^{N/2}g(I)$;
3) The matrix for inversion: $g(I)= (g(C_N^1))^{N/2} g(\sigma_h)$;
4) The matrix for mirror plane $\sigma_v$, which is parallel to the principal axis and contains $U_2$ one: $g(\sigma_v)=g(U_2)g(\sigma_h)$.

# Appendix B
## The factor-group analysis of the vibrational spectra for chiral and achiral SWCNTs

In permutation representation a symmetry element character [15] is equal to the number of those carbon atoms per one SWCNT cell, which are invariant or translated into equivalent positions (with respect to the translations of the GL) under this symmetry operation. In permutation representations of the armchair SWCNT only matrixes of identity element and mirror plane $\sigma_h$ have non-zero characters. In vector representation the character of $\sigma_h$ matrix is equal to 1, and only this element preserves all atomic positions or translates them into equivalent ones. So the number $Q_\alpha$ of the vibrational modes, which are spanned by the representation α can be expressed as:

$$Q_\alpha = \frac{1}{4N}\left(6N \cdot \chi_\alpha(E) + 2N \cdot \chi_\alpha(\sigma_h)\right) = \left(3 \cdot \chi_\alpha(E) + \chi_\alpha(\sigma_h)\right)/2, \qquad (8)$$

where character $\chi_\alpha(\sigma_h)$ of the IRREP α can be easily obtained with the use of formulas from Appendix A.

In permutation representations of the zigzag SWCNT only matrixes of identity element and the ones of $\sigma_v$ class have non-zero characters. Character of the latter ones is equal to 4, and the corresponding class includes $N/2$ elements. Therefore,

$$Q_\alpha = \left(3 \cdot \chi_\alpha(E) + \chi_\alpha(\sigma_v)\right)/2. \qquad (9)$$

As for chiral SWCNTs, only identity element has non-zero character in permutation representation. It is equal to $2N$. Thus the formula for the number $Q$ of the modes, which transform as representation α, can be written as

$$Q_\alpha = 3 \cdot \chi_\alpha(E). \qquad (10)$$

Using the formulas (8-10), one can easily obtain the results of factor-group analysis performed in [6] taking into account reported corrections [12].

# Appendix C
## The mode parity with respect to the mirror plane $\sigma_h$

In order to define whether the particular mode from Table 3 is even or odd with respect to the mirror $\sigma_h$, it is sufficient to consider the symmetrical coordinate [11], corresponding to the normal mode. It is so since symmetrical and normal coordinates are spanned by the same IRREP. Symmetrical coordinate of the GL mode of j-type can be expressed as: $s_k^j = p_j \exp(i\boldsymbol{kr})$, where $p_j$ is the form-factor (amplitude) function, which has the periodicity of the GL translations and is spanned by one-dimensional IRREP of the $G_k$ point group. Function $p_j$ determines the type of the GL mode (listed in the upper row of Table 3). The parity of functions $s_k^j$ with respect to $\sigma_h$ coincide with that of functions $p_j$, because the wave vector $\boldsymbol{k}$ belongs to the plane $\sigma_h$. Some of the $p_j$ functions are shown in Figure 3. These functions are related to the modes forming the G-band.

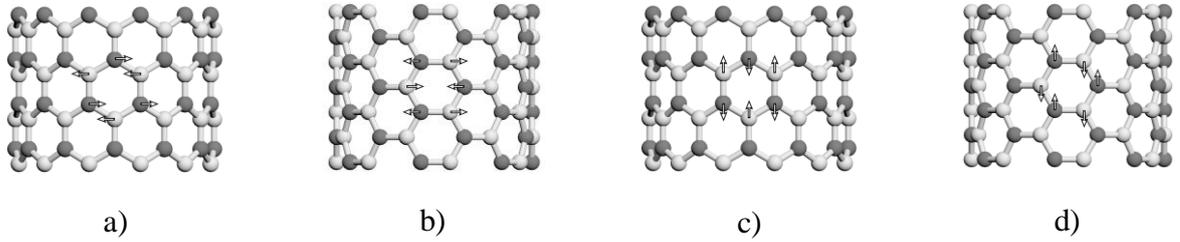

a)         b)         c)         d)

Figure 3. Atomic shifts corresponding to the functions $p_j$ (shown for one hexagon in each case) in the achiral SWCNTs. Functions of the iLO and iTO type are presented in panels (a, b) and (c, d) respectively. Functions shown in (b, c) are even with respect to $\sigma_h$, and the ones in (a, d) are odd.

**References**

1. L. A. Falkovsky, Phys. Lett. A **372**, 5189 (2008).
2. K. F. Mak, L. Jub, F. Wang, T. F. Heinz, Solid State Communications, **152**, 1341 (2012).
3. R. Saito, G. Dresselhaus, and M. S. Dresselhaus, *Physical Properties of Carbon Nanotubes*, Imperial College Press, London (1998).
4. M. S. Dresselhaus and P. C. Eklund, Adv. Phys. **49**, 705 (2000).
5. Ge. G. Samsonidze, R. Saito, A. Jorio, M. A. Pimenta, A. G. Souza Filho, A. Grüneis, G. Dresselhaus, and M. S. Dresselhaus, J. Nanosci. Nanotech. **3**, 431 (2003).
6. Ofir E. Alon, Phys. Rev. B **63**, 201403 (2001).
7. E. B. Barros, A. Jorio, Ge. G. Samsonidze, R. B. Capaz, A. G. Souza Filho, J. Mendes Filho, G. Dresselhaus, and M. S. Dresselhaus, Physics Reports **431**, 261 (2006).
8. J. L. Birman, *Light and Matter Ib: Theory of Crystal Space Groups and Infra-Red and Raman Lattice Processes of Insulating Crystals*, Springer (1974).
9. R. Saito, M. Hofmann, G. Dresselhaus, A. Jorio and M. S. Dresselhaus, Advances in Physics, **60:3**, 413 (2011).
10. L.D. Landau, E.M. Lifshitz, *Statistical Physics,* Vol. 5 (3rd ed.), Butterworth-Heinemann, Oxford (1980).
11. O. V. Kovalev, *Representations of the Crystallographic Space Groups: Irreducible Representations, Induced Representations, and Corepresentations*, translated from the Russian by G. C. Worthey, Gordon and Breach Science Publishers, Singapore (1993).
12. Y. Wang, B. Zhang, Q. Jin, B. Li, D. Ding, X. Cao, Spectrochimica Acta Part A **68**, 1149 - 1152 (2007).
13. L. A. Falkovsky, Journal of Experimental and Theoretical Physics **105**, 397-403 (2007).
14. P.T. Araujo, P.B.C. Pesce, M.S. Dresselhaus, K. Sato, R. Saito, A. Jorio, Physica E: Low-dimensional Systems and Nanostructures, **42**:5, 1251 (2010).
15. L.D. Landau, E.M. Lifshitz, *Quantum Mechanics: Non-Relativistic Theory*, Vol. 3 (3rd ed.), Pergamon Press, Oxford (1977).